\documentclass[11pt,a4paper]{article}
\usepackage{jheppub_kim}

\usepackage{pdflscape}
\usepackage{amsmath}
\usepackage{amssymb}
\usepackage{dcolumn}
\usepackage{bm}
\usepackage{color}
\usepackage{epsfig}
\usepackage{amsfonts}
\usepackage{graphicx}
\usepackage{subfigure}
\usepackage{dcolumn}

\newcommand{\be}{\begin{equation}}
\newcommand{\ee}{\end{equation}}
\newcommand{\bea}{\begin{eqnarray}}
\newcommand{\eea}{\end{eqnarray}}

\def\be{\begin{equation}}
\def\ee{\end{equation}}
\def\bea{\begin{eqnarray}}
\def\eea{\end{eqnarray}}

\begin{document}

\title{Effects of hyperscaling violation and dynamical exponents on heavy quark potential and jet quenching parameter}
\author[a]{M. Kioumarsipour}
\author[a]{J. Sadeghi}

\affiliation[a]{Sciences Faculty, Department of Physics, University of Mazandaran, Babolsar, Iran}

\emailAdd{m.kioumarsipour@stu.umz.ac.ir}
\emailAdd{pouriya@ipm.ir}

\abstract{\\
In this paper, we investigate the heavy quark potential and the jet quenching parameter in a system with Lifshitz and hyperscaling violation exponents, by using the AdS/CFT correspondence. It is shown that the heavy quark potential and the jet quenching parameter are dependent upon the nonrelativistic parameters. We show how the heavy quark potential changes with the hyperscaling violation parameter $ \theta $, dynamical parameter $ z $, temperature $ T $ and charge $ Q $. Increasing $ z $ and $ \theta $ lead to increasing and decreasing the potential respectively. The potential decreases and increases by increasing $ Q $ and $ T $. It is investigated how the jet quenching parameter changes with the hyperscaling violation parameter $ \theta $ and dynamical parameter $ z $. Also, we add some electromagnetic field and obtain its effect on the jet quenching parameter and see as $ z $ and $ \theta $ increasing this parameter decreases and the electric and magnetic fields affect differently on that.

}

\maketitle

\section{Introduction}
The interesting results of the heavy ion collision experiments in RHIC and LHC indicate that there is a new phase of matter so-called quark-gluon plasma (QGP) \cite{adam,shur,adox}. In order to investigate the properties of this matter, non-perturbative techniques are needed. Such non-perturbative technique is available by the AdS/CFT correspondence \cite{mal,wit,gub} that connects the supergravity theories that live in $d+1$-dimension anti-deSitter (AdS) spacetimes to the quantum field theories that live on the boundary $d$-dimension. Because of the conformal invariance and supersymmetry, the $ \mathcal{N}=4 $ Super-Yang Mills (SYM) dose not exactly equal to QCD at zero temperature. But at nonzero temperature, both theories describe strongly coupled non-Abelian plasmas. For the AdS geometry, the dual field theory considered will be conformal invariant. While the AdS geometry has the conformal invariance, the real QCD is not a conformal invariant theory, therefore the conformal invariance of $ AdS_{5} $ should be broken somehow. For this purpose, many efforts have been done, for example the hard-wall \cite{pol,h1,h2,h3,h4} and soft-wall approximations \cite{h5,h6,h7,h8} that each one suffered of some problems. Recently, geometries have been considered that are not asymptotically AdS, therefore their dual field theories are not conformally invariant, but they are still scale invariant. One of the most important example is Lifshitz fixed point that has an anisotropic scale symmetry. Such systems are scaled anisotropically in the time direction by a dynamical exponent $ z $ and are spatially isotropic. These systems are invariant under the following scale transformations,
\begin{equation}
(t,\vec{x})\rightarrow (\lambda^{z}t,\lambda\vec{x}).
\end{equation}
If $ z=1 $, theory has a relativistic scale invariance and in the case $ z\neq 1 $, theory is non-relativistic because it is scaled differently in the time and space directions. In order to have such theories, the Einstein gravity must include the Abelian gauge fields \cite{lif1,lif2}. If theory of gravity couples with a scalar field also, the full class of metrics that exhibit scaling properties, will emerge \cite{lif3,lif4,lif5,lif6,lif7,lif8,lif9}. Considering both these fields, a metric will be achieved as the following form \cite{alii},
\begin{equation}
ds^{2}=r^{-2\theta /d}\left( -r^{2z}dt^{2}+\dfrac{dr^{2}}{r^{2}}+r^{2}d{\bf x}^{2}\right).\label{tt} 
\end{equation}
The above equation is spatially homogeneous and not invariant under scaling but represents the following scale transformations,
\begin{equation}
t\rightarrow \lambda^{z}t,~~~~r\rightarrow \lambda^{-1}r,~~~~x_{i}\rightarrow \lambda x_{i},~~~~ds_{d+2}\rightarrow \lambda^{\theta/2}ds_{d+2}.
\end{equation}
An interesting property of such metrics is that they break directly conformal invariance, and therefore they can be appropriate candidates to study QCD. In general, the metric (\ref{tt}) can not provide a good description in all range of $ r $ and one should follow an \textit{effective} holographic approach in which the dual theory lives on a finite $ r $ slice \cite{null}. Note that these theories are intrinsically non-relativistic, so one can consider them as toy models from the holography viewpoint. A useful review on the holographic modelling of field theories with Lifshitz symmetry was presented in \cite{tay}. In this review, gravity duals for non-relativistic field theories, the holographic dictionary for Lifshitz backgrounds, the relationship between bulk fields and boundary operators, and operator correlation functions was studied.

Since one of the experimental signatures of QGP formation is melting of heavy mesons, a heavy quark potential and screening radius are important in the study of QGP \cite{mat}. The first calculation of the heavy quark potential has been done by Maldacena for $ \mathcal{N}=4 $ SYM \cite{mald}. He showed that the energy indicates a purely Coulombian behavior for the $ AdS_{5} $ space that is agreement with a conformal gauge theory. After Maldacena, several times the heavy quark potential was calculated by using the AdS/CFT correspondence. For example, the heavy quark potential was calculated at finite temperature in $ \mathcal{N}=4 $ supersymmetric gauge theory \cite{rey,bran}, at a curved spacetime having an extra dimension \cite{kin}, as a function of shear viscosity in a strongly coupled field theory dual to five-dimensional Gauss-Bonnet gravity \cite{noro}, at the presence of higher derivative corrections \cite{fada}, in a strongly coupled non-conformal plasma with anisotropy \cite{enr}, in a $ D $-instanton background \cite{zi}, in the Lifshitz backgrounds with hyperscaling violation \cite{zii} and in a deformed $ AdS $ background \cite{ziii}. The velocity dependence quark-antiquark potential in which a dipole moves through the plasma with an arbitrary velocity was computed in $ \mathcal{N}=4 $ SU(N) Yang-Mills theory \cite{som} and in noncommutative Yang-Mills theory \cite{somd}. Also the quark-monopole potential for $ \mathcal{N}=4 $ super Yang-Mills was obtained in \cite{jos} and it was resulted that the quark-antiquark and monopole-antimonopole potentials are more than the strength of the quark-monople potential. Also, authors in \cite{jun} calculated the screening length of meson with different angular momentum and dependency of the temperature of the screening length, and they incorporated the angular momentum of the mesonic states by rotating the metric background.

Another interesting property of the strongly-coupled plasma is jet quenching parameter that is related to momentum fluctuation  \^{q}$\propto \langle \vec{p~}_{\bot}^{2} \rangle $. The first calculation of the jet quenching parameter for $ \mathcal{N}=4 $ SYM was done in \cite{liu}. Since the knowledge about the jet quenching parameter increases our insight about the quark gluon plasma, studying of this parameter is interesting. After \cite{liu}, the AdS/CFT correspondence is used widely to calculate this quantity. For example, the effects of charge and finite 't Hooft coupling correction on the jet quenching parameter was calculated \cite{fad}. Also, the jet quenching parameter was investigated in the $ \mathcal{N}=2 $ thermal plasma \cite{has}, in the anisotropic systems \cite{wan,che}, in a medium with chemical potential \cite{fen} and in power-law Maxwell field system \cite{khan}.

Motivated by these considerations, in this paper we will study the heavy quark potential and the jet quenching parameter in a hyperscaling background.
We are going to investigate the effects of all the parameters of the theory such as charge, dynamical exponent $ z $ and hyperscaling violation exponent $ \theta $ on these quantities, and show how they affect our results. We will see that the heavy quark potential and the jet quenching parameter can change by changing the values of these parameters, that is to say, these quantities depend on the nonrelativistic parameters. Moreover, we will compare the results with those of conformal case and experimental data. 

This paper is organized as follows. In section \ref{sec2}, we investigate the heavy quark-antiquark potential in the hyperscaling violation background and analyze the effects of charge $ Q $, temperature $ T $, dynamical exponent $ z $ and hyperscaling violation $ \theta $ on it. In section \ref{sec3}, we calculate the jet quenching parameter and then in section \ref{sec4}, we add a constant electromagnetic field to the system and study its effect on the jet quenching parameter. Finally, In section \ref{sec5} we summarize our results.

\section{Heavy quark potential}\label{sec2}
Now, we want to investigate the heavy quark-antiquark potential in the hyperscaling violation metric background. For this purpose, we consider the black brane solution of the Einstein-Maxwell-Dilaton system introduced in \cite{alii},
\begin{eqnarray}
ds^{2}&=&r^{2\alpha}\left( -r^{2z}f(r)dt^{2}+\dfrac{dr^{2}}{r^{2}f(r)}+r^{2}d{\bf x}^{2}\right), ~~~~\alpha:=-\theta /d, \label{ads}\nonumber\\
f(r)&=&1-\dfrac{m}{r^{z+d-\theta}}+\dfrac{Q^{2}}{r^{2(z+d-\theta-1)}},
\end{eqnarray}
where $ z $ is dynamical exponent and $ \theta $ is hyperscaling violation exponent, and $ m $ and $ Q $ are related to the mass and charge of the black hole respectively. By setting $ f=0 $, one  can obtain the radius of horizon $ r_{h} $ as following,
\begin{equation}
r_{h}^{2(d+z-\theta -1)}-mr_{h}^{d+z-\theta -2}+Q^{2}=0.
\end{equation}
The Hawking temperature $ T $ is given by,
\begin{equation}
T=\dfrac{r_{h}^{z}(d+z-\theta)}{4\pi}\left( 1-\dfrac{(d+z-\theta -2)Q^{2}}{d+z-\theta}r_{h}^{2(\theta -d-z+1)}\right).\label{TT} 
\end{equation}

It is known, to have a physically sensible dual field theory, the null energy condition (NEC) should be satisfied at least from the gravity side. Therefore, one can write $ T_{\mu\nu}N^{\mu}N^{\nu}\geq 0 $ for an arbitrary null vector $ N^{\mu} $ \cite{null,null1}. Applying the NEC on (\ref{ads}) leads to the following condition,
\begin{equation}
(d-\theta)(d(z-1)-\theta)\geq 0.\label{a} 
\end{equation}
By using this condition one can obtain the allowed values for $ (z,\theta) $ so that the given gravity dual will be consistent. If $ z=1 $, the theory is Lorentz invariant and (\ref{a}) implies $ \theta \leq 0 $ or $ \theta\geq d $. In the range of $ \theta > d $ there is some instabilities in the gravity side and this range is just allowed based on the NEC \cite{null}. Therefore, we only consider the range of $ \theta\leq d $. For $ \theta =0 $, that exhibits a scale invariant theory, one can reachieve the known result $ z\geq 1 $.

In the case $ \theta\leq d $, we can obtain the following dynamical exponent from the condition (\ref{a}),
\begin{equation}
z\geq 1+\frac{\theta}{d}. \label{b} 
\end{equation}
If we consider $ \theta =d, d-1 $ and $ d/2 $, therefore we find $ z\geq 2, 2-1/d $ and $ 3/2 $ respectively. 

The range of $ d-1\leq\theta\leq d $ displays novel phases that show new violations of the area law that interpolate between logarithmic and linear behaviors \cite{null}. The case $ \theta=d $ points out an extensive violation of the entanglement entropy. In theories with $ \theta=d-1 $ the holographic entanglement entropy exhibits a logarithmic violation of the area law which indicates the model must have a Fermi surface \cite{fermi,fermi1}.

Before we proceed further, it is important to mention a point. As mentioned, the geometries with nontrivial hyperscaling violation could be thought of as an \textit{effective} holographic description of quantum field theory which lives on a finite $ r $ slice. So, the metric background (\ref{ads}) can be considered as an \textit{effective} theory which is taken to live at finite $ r $ of order $ r_{F} $. Therefore, the metric (\ref{ads}) provides a good description of the dual field theory which lives on a background spacetime identified with a surface constant of $ r $ in (\ref{ads}). In other words, in our analysis we put our \textit{boundary} QFT at an effective UV cut off.

For obtaining the heavy quark potential one can calculate the expectation value of the Wilson loop from the holography point of view,
\begin{equation}
W(\mathcal{C})=\dfrac{1}{N} Tr P exp\left( i \oint_{\mathcal{C}} A_{\mu} dx^{\mu}\right)
\end{equation}
where $ \mathcal{C} $ is a rectangular loop along the time $ \mathcal{T} $ and spatial extension $ L $ in the spacetime, the trace is over the fundamental representation of SU(N) group, $ A_{\mu} $ is the gauge potential and $ P $ enforces the path ordering.

In the limit $ \mathcal{T}\rightarrow \infty $, one can obtain the heavy quark potential from the expectation value of Wilson loop,
\begin{equation}
 <W(\mathcal{C})> \sim e^{-i\mathcal{T}V_{q\overline{q}}}.
\end{equation}
Also, according to the AdS/CFT duality, the expectation value of $ W(\mathcal{C}) $ is given by,
\begin{equation}
 <W(\mathcal{C})> \sim e^{-iS(\mathcal{C})},
\end{equation}
where $ S_{\mathcal{C}} $ is the regularized action. Hence, the heavy quark potential can be obtained by,
\begin{equation}
V_{q\overline{q}}(L)=\dfrac{S_{\mathcal{C}}}{\mathcal{T}}.
\end{equation}

Here, we consider the heavy quark-antiquark potential in the hyperscaling violation metric background and start with the Nambu-Goto action,
\begin{equation}
S=-\dfrac{1}{2\pi\alpha^{\prime}}\int d\tau d\sigma \sqrt{-det g_{\alpha\beta}},\label{NG}
\end{equation}
where $ g $ is the determinant of the induced metric of the string worldsheet,
\begin{equation}
g_{\alpha\beta}=G_{\mu\nu}\partial_{\alpha}X^{\mu} \partial_{\beta}X^{\nu}.
\end{equation} 
By considering $ X^{\mu}=(t,x,0,0,r(x)) $ for the static string coordinates and the static gauge $ \sigma=x, \tau=t $ and $ r=r(x) $, we obtain the induced metric of the fundamental string as,
\begin{equation}
g_{\alpha\beta}=r^{2(\alpha +1)}\left( 
\begin{array}{ccc}
-r^{2(z-1)}f(r) & 0 & \\
0 &1+\dfrac{r^{\prime 2}}{r^{4}f(r)}  &
\end{array} \right).
\end{equation}
where $ \prime $ is a derivative with respect to $ \sigma=x $. Then we can find the Euclidean version of the Nambu-Goto action (\ref{NG}) as,
\begin{equation}
S=\dfrac{\mathcal{T}}{2\pi\alpha^{\prime}}\int dx \sqrt{r^{\beta}\left( f+\dfrac{r^{\prime 2}}{r^{4}} \right) },~~~~~\beta:= 4\alpha +2z+2. \label{9}
\end{equation}
It is obvious that the Lagrangian,
\begin{equation}
\mathcal{L}=\sqrt{r^{\beta}\left( f+\dfrac{r^{\prime 2}}{r^{4}} \right) },
\end{equation}
does not depend on $ x $, so the corresponding Hamiltonian $ H $ is a constant of motion,
\begin{equation}
H=\dfrac{\partial \mathcal{L}}{\partial r^{\prime }}r^{\prime}-\mathcal{L}=constant.
\end{equation}
The boundary condition shows that the string configuration has a maximum at $ x=0 $, i.e. $ r(0)=r_{c} $, so that $ r^{\prime}_{c}=0 $, then we obtain,
\begin{equation}
-\dfrac{r^{\beta}f}{\sqrt{r^{\beta}\left( f+\dfrac{r^{\prime 2}}{r^{4}} \right) }}=-\sqrt{r_{c}^{\beta}f_{c}}.
\end{equation}
Using this condition we find,
\begin{equation}
r^{\prime}=\sqrt{r^{4}f\left( \dfrac{r^{\beta}f}{r_{c}^{\beta}f_{c}}-1 \right) },\label{8}
\end{equation}
where $ f_{c}=f(r_{c}) $. Then one can derive the following separation length $ L $ of the quark-antiquark by integrating (\ref{8}),
\begin{equation}
L=2\int dr \sqrt{\dfrac{r_{c}^{\beta}f_{c}}{r^{4}f\left( r^{\beta}f-r_{c}^{\beta}f_{c} \right) }}.\label{LLA}
\end{equation}
In order to find the action of the heavy quark pair, one should substitute (\ref{8}) into (\ref{9}),
\begin{equation}
S=\dfrac{\mathcal{T} }{\pi\alpha^{\prime}}\int dr \sqrt{\dfrac{r^{2\beta}f}{r^{4}\left( r^{\beta}f-r_{c}^{\beta}f_{c} \right) }}. \label{11}
\end{equation}

Now, one should subtract the self energy of the two quarks to avoid of the divergence of the action (\ref{11}),
\begin{equation}
S_{0}=\dfrac{\mathcal{T} }{\pi\alpha^{\prime}}\int dr r^{\gamma},~~~~~~\gamma:= 2\alpha +z-1.
\end{equation}
Finally, one can obtain the heavy quark potential in the hyperscaling violation metric background as following,
\begin{equation}
V_{q\bar{q}}(L)=\dfrac{1}{\pi\alpha^{\prime}}\int_{r_{c}}^{r_{F}}  dr r^{\gamma} \left( r^{\gamma+2}\sqrt{\dfrac{f}{r^{\beta}f-r_{c}^{\beta}f_{c}}}-1 \right) -\dfrac{1}{\pi\alpha^{\prime}}\int _{r_{h}}^{r_{c}} dr r^{\gamma}.\label{221}
\end{equation}
\begin{figure}
\begin{center}$
\begin{array}{cc}
\includegraphics[width=74 mm]{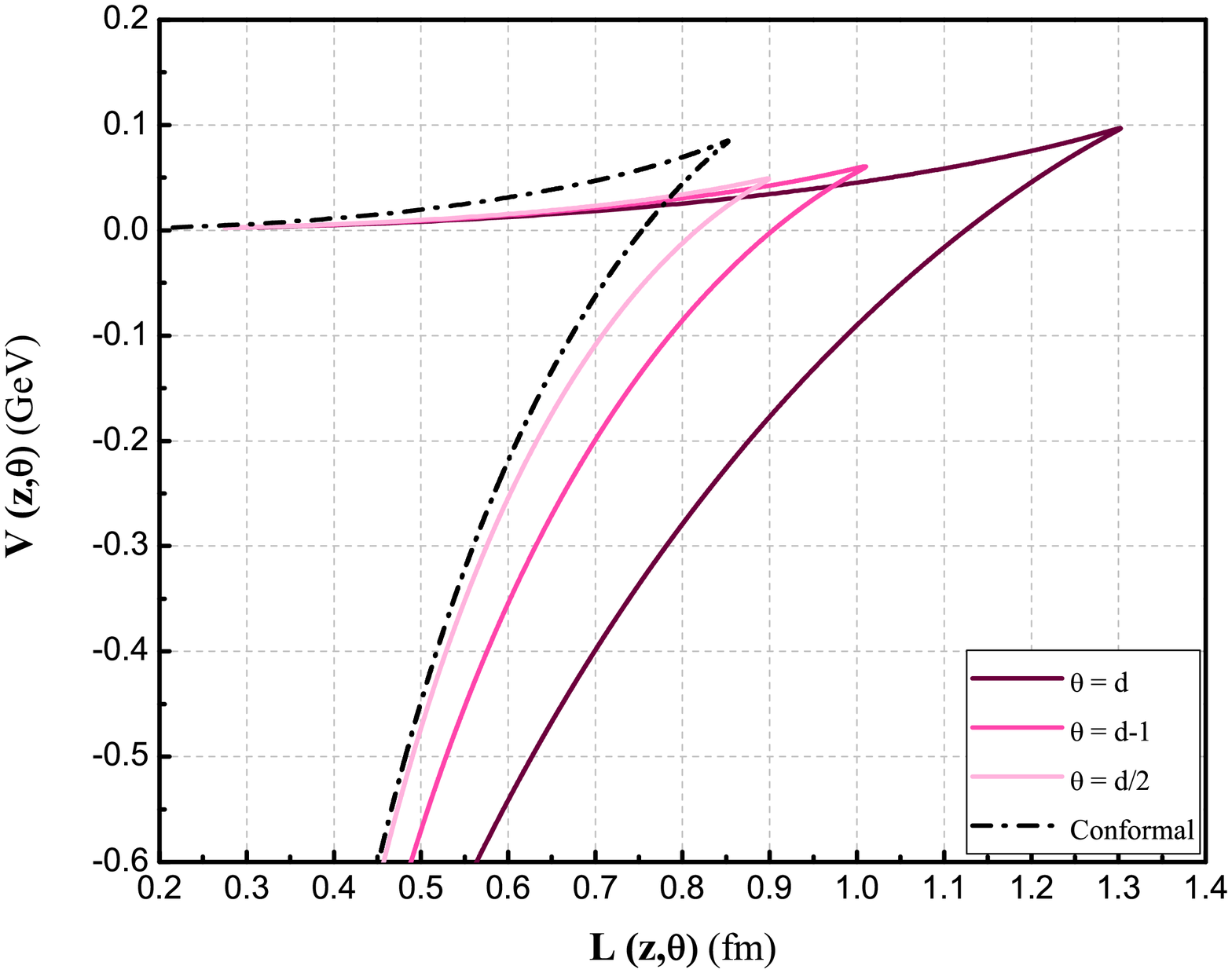}
\includegraphics[width=74 mm]{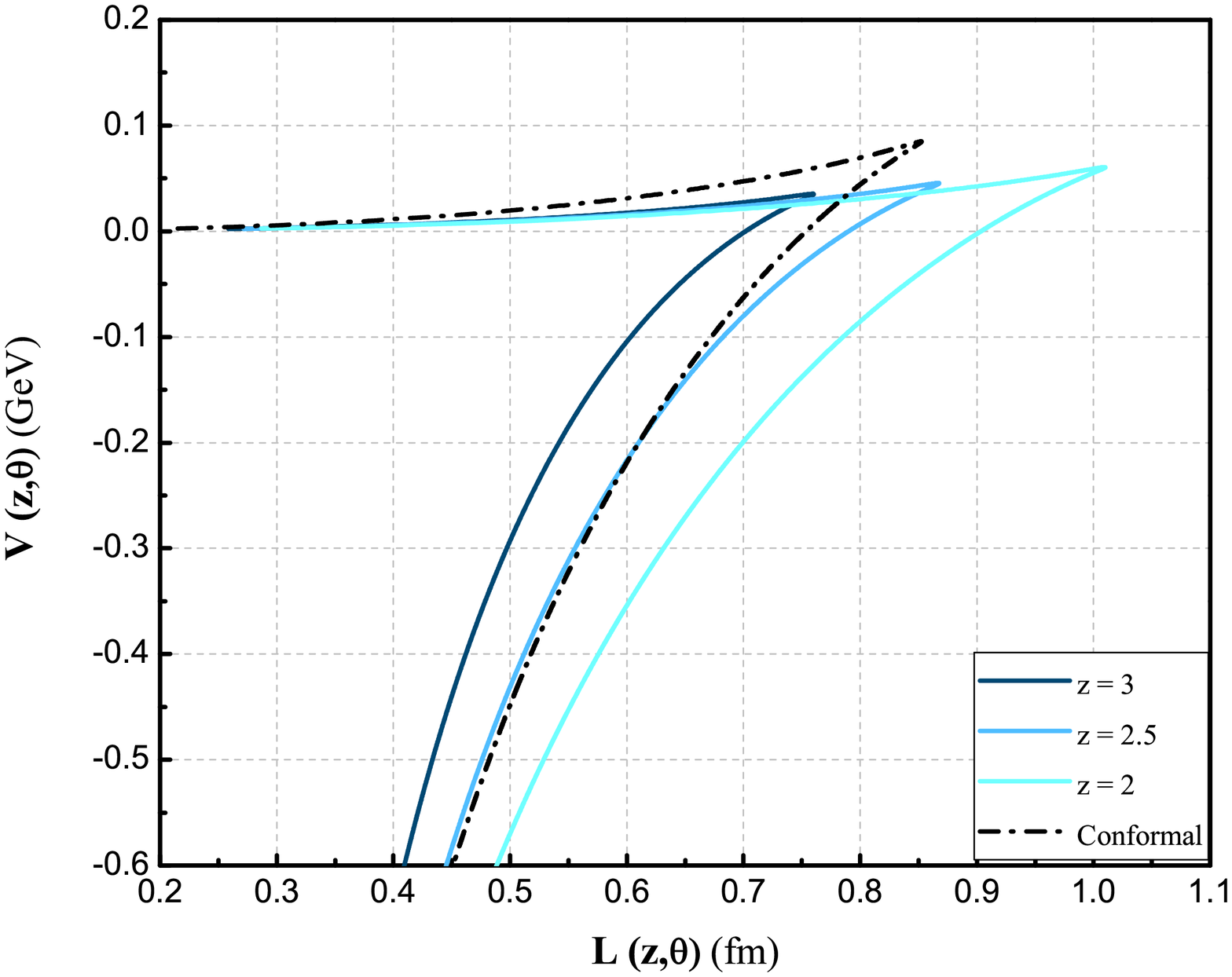}
\end{array}$
\end{center}
\caption{Left: the heavy quark potential versus the quark-antiquark separation in different $ \theta $ at $ z=2 $ and $ Q=1 $. Right: the heavy quark potential versus the quark-antiquark separation in different $ z $ at $ \theta=d-1 $ and $ Q=1 $.}
\label{ww}
\end{figure}

\begin{figure}
\begin{center}$
\begin{array}{cc}
\includegraphics[width=74 mm]{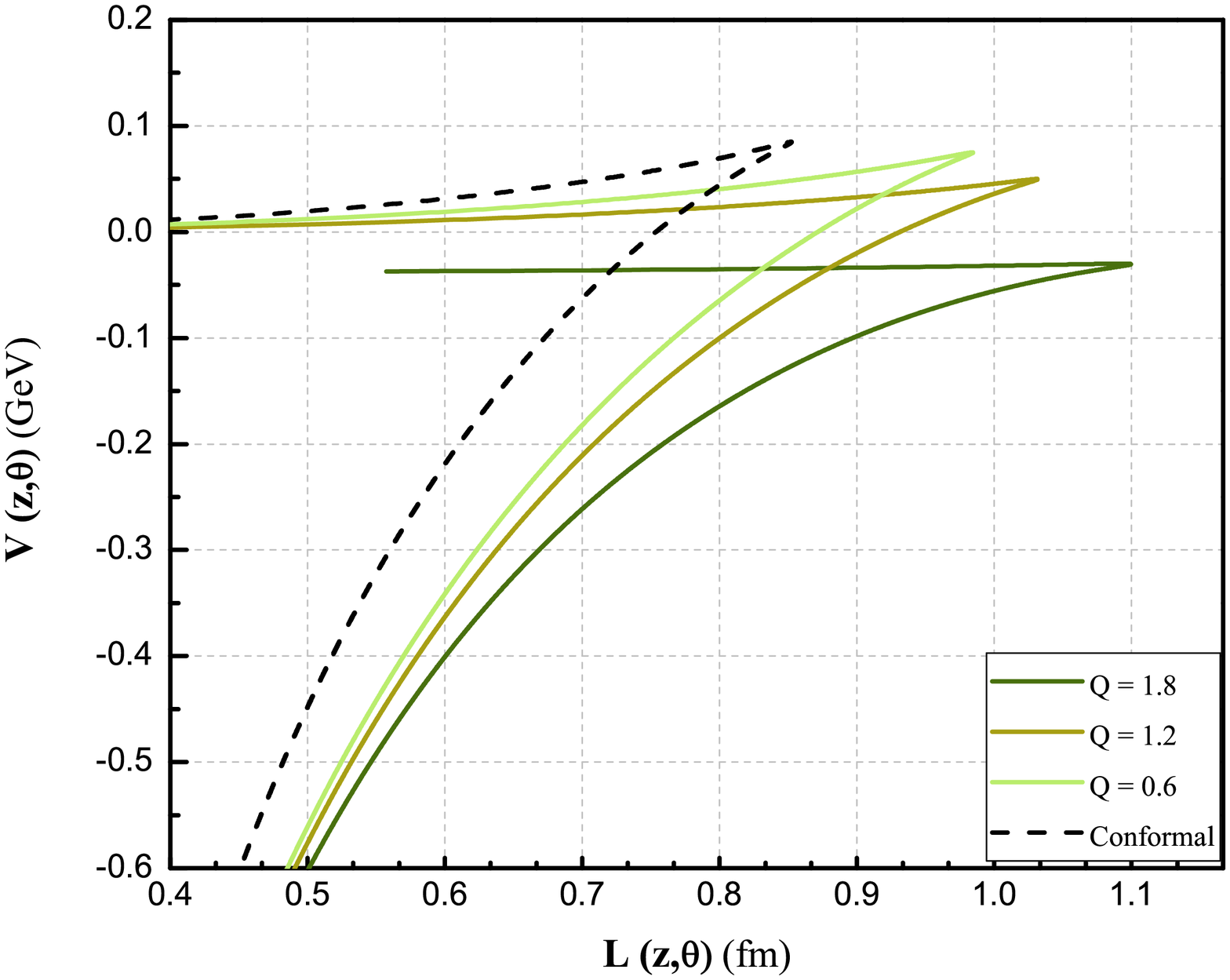}
\includegraphics[width=74 mm]{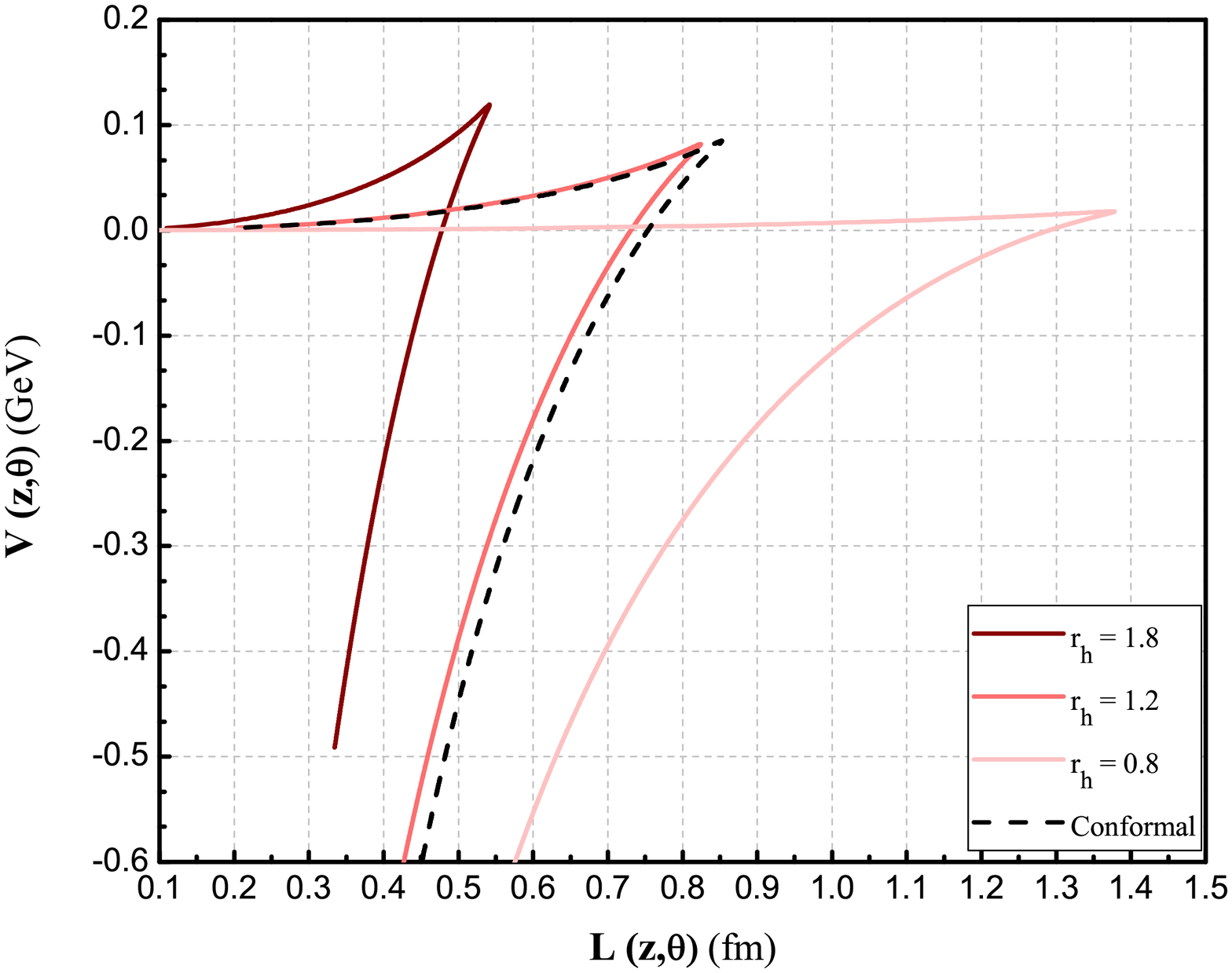}
\end{array}$
\end{center}
\caption{Left:  the heavy quark potential versus the quark-antiquark separation in different $ Q $ at $ \theta=d-1 $ and $ z=2 $. Right:  the heavy quark potential versus the quark-antiquark separation in different $ r_{h} $ at $ \theta=d-1 $, $ z=2 $ and $ Q=1 $.}
\label{w}
\end{figure}
For the convenience of numerical calculation we can make a transformation $ u=r_{c}/r $, so that the horizon of the black brane is located at $ u=r_{c}/r_{h} $ and the boundary of the spacetime is placed in the limit, $ u\rightarrow 0 $. With this definition, the equations (\ref{LLA}) and (\ref{221}) become,
\begin{eqnarray}
L&\!=\!& 2\int _{\epsilon}^{1} \dfrac{du}{u} \sqrt{\dfrac{u^{\beta+2}f(r_{c})}{r_{c}^{2}f(u)\left[f(u)-u^{\beta}f(r_{c}) \right] }},\\
V_{q\bar{q}}(L)&\!=&\!\dfrac{1}{\pi\alpha^{\prime}}\left ( \int _{\epsilon}^{1}\dfrac{du}{u}\left (\dfrac{r_{c}}{u}\right )^{\gamma+1}\left (\sqrt{\dfrac{f(u)}{f(u)-u^{\beta}f(r_{c})}}-1\right )-\int _{1}^{\frac{r_{c}}{r_{h}}} \dfrac{du}{u}\left (\dfrac{r_{c}}{u}\right )^{\gamma+1}\right ),~~~
\end{eqnarray}
where $ \epsilon $ is the so called \textit{effective} UV cutoff.

Now, we want to investigate the effects of the dynamical exponent and the hyperscaling violation exponent on the heavy quark potential and on the interquark distance. It is crystal clear that in the different values $ z $ and $ \theta $ the dependency of the quark potential and interquark distance to $ r $ is different and complicated and we can not solve them analytically, therefore, inevitably we should resort to numerical methods. In figure \ref{ww}, we plot numerically the heavy quark potential $ V(\theta,z) $ as a function of the interquark distance $ L(\theta,z) $ at a fixed temperature ($ h=1 $). The left plot shows the influence of the hyperscaling violation factor on the heavy quark potential at fixed dynamical exponent $ z $. As we see, increasing $ \theta $ leads to decreasing the heavy quark potential. The influence of the dynamical exponent on the potential is shown in the right plot. It is obvious that the heavy quark potential increases by increasing $ z $ at the fixed $ \theta=d-1 $. In fact, by increasing $ z $ and $ \theta $, the potential behaves differentially. These results are in agreement with results in \cite{zii}. In figure \ref{w}, we want to study the effects of the charge and temperature on the heavy quark potential and on the interquark distance, hence, we illustrate the heavy quark potential $ V(\theta,z) $ versus the distance $ L(\theta,z) $ in different values of the charge and temperature. From the left plot we can see that increasing charge leads to decreasing the potential. As we see, by increasing temperature $ T $ (or increasing $ r_{h} $) the heavy quark potential increases, indeed $ T $ is a ascending function of $ r_{h} $ at $ z=2 $, $ \theta=d-1 $ and $ Q=1 $. This result is in contradiction with \cite{zii}, since their temperature is a decreasing function of $ r_{h} $ at the special $ z=1.6 $.

Comparing plots we can see that the form of the potential is similar to the conformal case and the closest state is $ z=2 $, $ \theta=d-1 $, $ Q=1 $ and $ r_{h}=1.2 $.

\section{Jet quenching}\label{sec3}
In this section, we evaluate the jet quenching parameter in the hyperscaling violation metric background. This quantity provides a measure of the strenght of an energetic parton  which interact with QGP.

The jet quenching parameter is related to the expectation value of the Wilson loop in the adjoint representation,
\begin{equation}
\langle W^{A}(\mathcal{C})\rangle  \approx exp \left( -\dfrac{1}{4\sqrt{2}}\hat{q}L^{-}L^{2}\right),
\end{equation}
where the contour $ \mathcal{C} $ is a rectangular Wilson loop with the small extension $ L $ along a transverse direction and the large extension $ L^{-}$ along the light-cone that was formed by a pair of quark-antiquark ($q\bar{q}$) \cite{liu}. According to the AdS/CFT correspondence, one can calculate the thermal expectation value of a Wilson loop in the fundamental representation as,
\begin{equation}
\langle W^{F}(\mathcal{C})\rangle = exp[-S_{I}(\mathcal{C})],
\end{equation}
where $ S_{I}=S-S_{0} $. Here $ S $ is the action of the extremal surface in the AdS spacetime whose boundary is the contour $ \mathcal{C} $ enclosed the world-sheet surface of the string and $ S^{0} $ is self-energy of the quark and antiquark. Using the relation $ \langle W^{A}(\mathcal{C})\rangle \simeq \langle W^{F}(\mathcal{C})\rangle^{2} $, the jet quenching parameter becomes,
\begin{equation}
\hat{q}=8\sqrt{2}\dfrac{S_{I}}{L^{-}L^{2}}.\label{5}
\end{equation}

By introducing the light-cone coordinate $ x^{\mu}=(r,x^{+},x^{-},x_{2},x_{3}) $, we rewrite the hyperscaling metric background (\ref{ads}) as,
\begin{eqnarray}
ds^{2}= r^{2\alpha +2}& \Bigg( & \dfrac{1-r^{2z-2}f}{2}\left( (dx^{+})^{2}+(dx^{-})^{2}\right) -\left( 1+r^{2z-2}f\right) dx^{+}dx^{-} \nonumber\\ &&+ dx^{2}_{2}+dx^{2}_{3}+\dfrac{dr^{2}}{r^{4}f} \Bigg).
\end{eqnarray}
Assuming that the quark-antiquark pair are situated at $ x_{2}=\pm \frac{L}{2} $ on $ x^{+}=x_{3}=const $ plane and choosing the static gauge $ \tau=x^{-}(0 \leq x^{-} \leq L^{-}) $, $ \sigma=x_{2}(-\frac{L}{2} \leq x_{2} \leq \frac{L}{2}) $, we obtain the Nambu-Goto action as following, 
\begin{eqnarray}
S&=&\dfrac{1}{2\pi \alpha^{\prime}}\int d\tau d\sigma \sqrt{detg_{\alpha\beta}}\nonumber\\ &=&\dfrac{L^{-}}{\sqrt{2}\pi \alpha^{\prime }}\int_{0}^{L/2} d\sigma r^{2(\alpha +1)}\sqrt{\left(1- r^{2(z-1)}f\right) \left( 1+\dfrac{r^{\prime 2}}{r^{4}f}\right) },~~~~~r^{\prime}=\dfrac{dr}{d\sigma}. \label{2} 
\end{eqnarray}
The above equation is calculated  by substituting the induced metric of the string in the action. The string configuration is specified by $ r=r(\sigma) $, since in the limit $ L^{-}\gg L $ the worldsheet is translationally invariant along the $ \tau $ direction. The integrand in (\ref{2}) is independent of $ \sigma $, thereupon, there is a conserved quantity, then we obtain the equation of motion for $ r(\sigma) $ as,
\begin{equation}
r^{\prime 2}=\dfrac{r^{4}f}{2H^{2}}\left( r^{4(\alpha +1)}\left(1- r^{2(z-1)}f \right) -2H^{2} \right) ,~~~~\label{3}
\end{equation}
where $ H $ is the constant of motion. Using the boundary conditions $ r(\pm\frac{L}{2})=\infty $ and $ r^{\prime}=0 $, we can determine the turning point $ r_{min} $ ($ \geq r_{h} $) by solving (\ref{3}). Since we consider the case with small $ H $, we see that the factor $  r^{4(\alpha +1)}\left( 1-r^{2(z-1)}f \right) -2H^{2} $ is always positive near the black hole horizon and negative near the boundary. The turning point $ r_{min} $ is obtained by solving the following equation,
\begin{equation}
 1-r_{min}^{2(z-1)}f(r_{min})=0 .
\end{equation}
Substituting (\ref{3}) and $ d\sigma =dr/r^{\prime} $ in the Nambu-Goto action (\ref{2}), we find,
\begin{equation}
S=\dfrac{L^{-}}{\sqrt{2}\pi \alpha^{\prime}}\int dr\sqrt{\dfrac{r^{4\alpha}\left(1- r^{2(z-1)}f \right) }{f}}\left( 1-\dfrac{2H^{2}}{r^{4(\alpha +1)}\left(1- r^{2(z-1)}f \right) } \right)^{-\frac{1}{2}}.\label{4}
\end{equation}
Since $ H $ is very small, we can expand (\ref{4}) to leading order of $ H^{2} $,
\begin{equation}
S=\dfrac{L^{-}}{\sqrt{2}\pi \alpha^{\prime}}\int dr\sqrt{\dfrac{r^{4\alpha}\left( 1-r^{2(z-1)}f \right) }{f}}\left( 1+\dfrac{H^{2}}{r^{4(\alpha +1)}\left( 1-r^{2(z-1)}f \right) } \right).
\end{equation}
To avoid the divergence this action which results from the contribution of mass of two quarks, the self energy of two quark should be subtracted from the above action. The self energy is given by,
\begin{eqnarray}
S_{0}&=&\dfrac{2L^{-}}{2\pi \alpha^{\prime}}\int d\sigma \sqrt{g_{--}g_{rr}}\nonumber\\
&=&\dfrac{L^{-}}{\sqrt{2}\pi \alpha^{\prime}}\int dr \sqrt{\dfrac{r^{4\alpha}}{f}\left(1- r^{2(z-1)}f \right) }.
\end{eqnarray}
Then the regularized action is given by,
\begin{equation}
S_{I}=S-S_{0}=\dfrac{L^{-}H^{2}}{\sqrt{2}\pi \alpha^{\prime}}\int dr \Big( r^{4(\alpha +2)}f \left( 1-r^{2(z-1)}f \right)  \Big)^{-\frac{1}{2}}.\label{6}
\end{equation}

Also, we can calculate the distance between two quarks by integrating of the equation (\ref{3}) as,
\begin{equation}
\dfrac{L}{2}=\int  \dfrac{\sqrt{2}Hdr}{\sqrt{r^{4(\alpha +2)}f\left(1- r^{2(z-1)}f \right) }}\left( 1-\dfrac{2H^{2}}{r^{4(\alpha +1)}\left( 1-r^{2(z-1)}f\right) } \right)^{-\frac{1}{2}}.
\end{equation}
Since we considered the case with small $ H $, we can expand the above equation and obtain the below relation,
\begin{equation}
\dfrac{L}{2H}\simeq \int_{0}^{L/2} dr \left( \dfrac{1}{2}r^{4(\alpha +1)}f\left(1- r^{2(z-1)}f \right) \right)^{-\frac{1}{2}}.\label{7}
\end{equation}
Substituting (\ref{6}) and (\ref{7}) into (\ref{5}), we finally obtain the jet quenching parameter for the hyperscaling metric background as following,
\begin{equation}
\hat{q}=\dfrac{1}{\pi\alpha^{\prime}}\left( \int_{r_{h}}^{r_{F}} \dfrac{dr}{\sqrt{r^{4(\alpha +2)}f(r)\left(1- r^{2(z-1)}f(r) \right) }} \right)^{-1}.
\end{equation}
We define a transformation $ u=r_{h}/r $ for the convenience of numerical calculation. Whit this definition, the above equation become,
\begin{equation}
\hat{q}=\dfrac{1}{\pi\alpha^{\prime}}\Bigg[ \int _{\epsilon}^{1}\dfrac{du}{u^{2}}\dfrac{r_{h}}{\sqrt{\left (\dfrac{r_{h}}{u}\right )^{4(\alpha +2)}f(u)\left(1- \left (\dfrac{r_{h}}{u}\right )^{2(z-1)}f(u) \right) }} \Bigg]^{-1},
\end{equation}
where $ \epsilon $ is the UV cutoff.

\begin{figure}
\begin{center}$
\begin{array}{cc}
\includegraphics[width=74 mm]{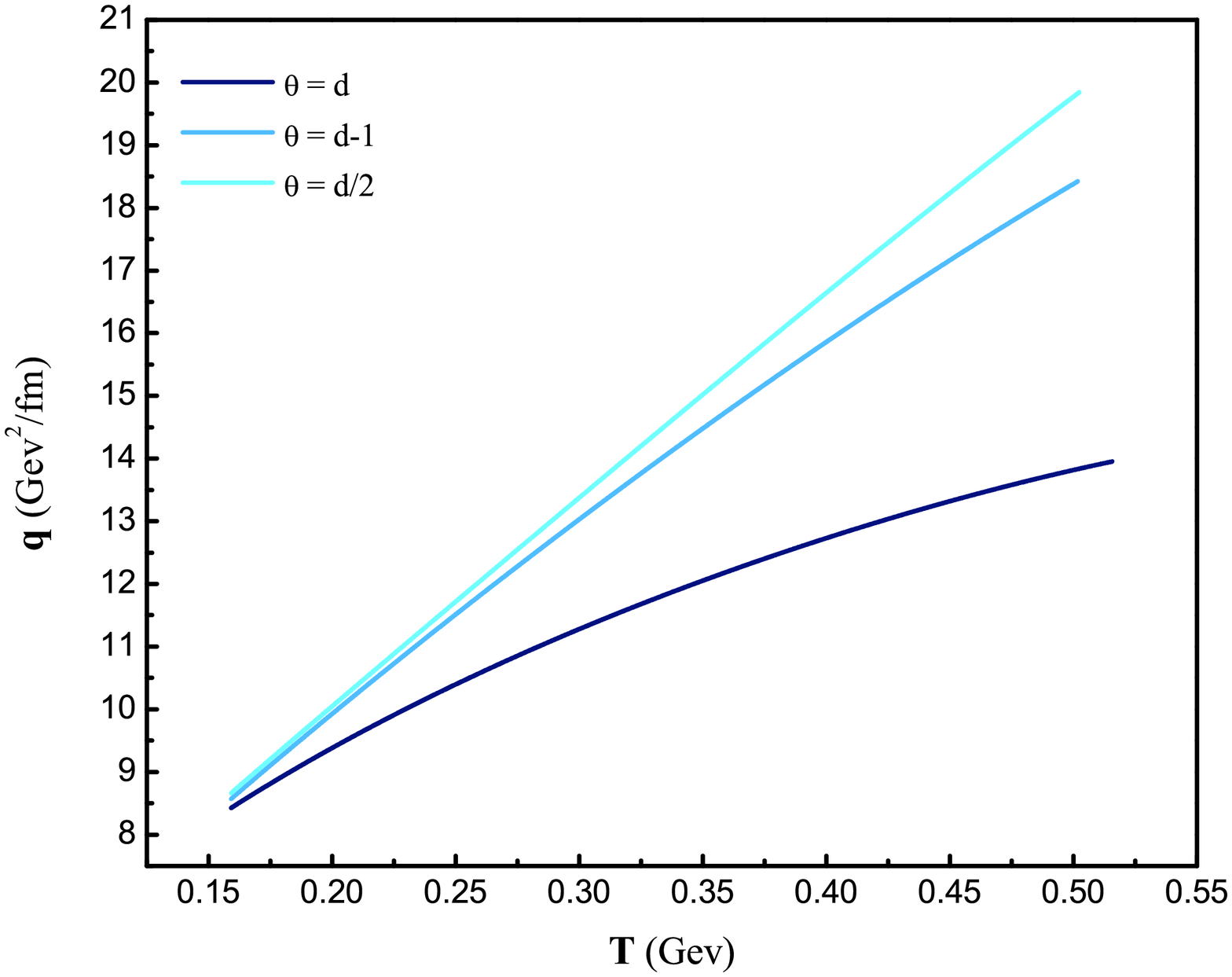}
\includegraphics[width=74 mm]{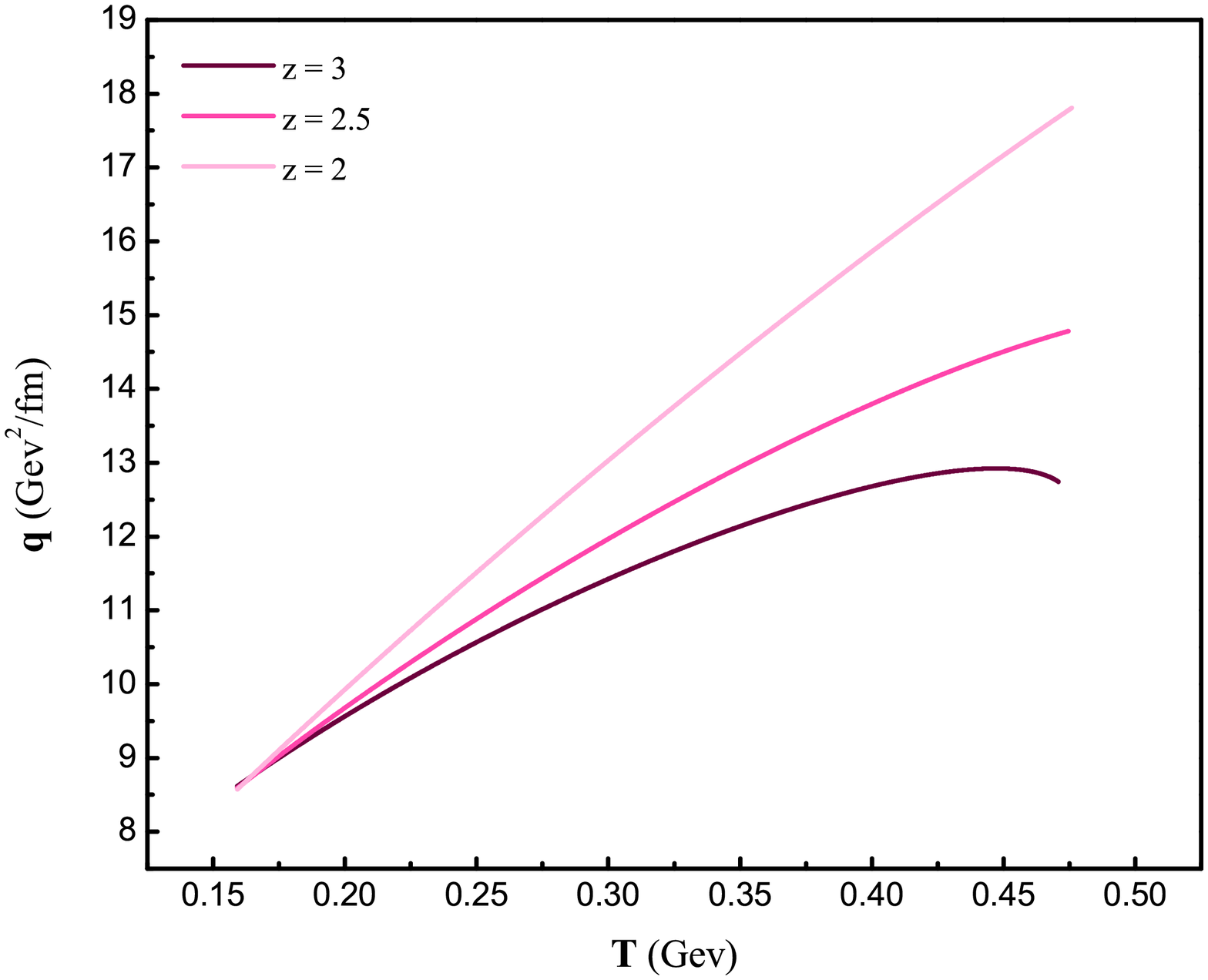}
\end{array}$
\end{center}
\caption{Left: jet-quenching versus temperature in different $ \theta $ at $ z=2 $ and $ Q=1 $. Right: jet-quenching versus temperature in different $ z $ at $ \theta=d-1 $ and $ Q=1 $.}
\label{www}
\end{figure}

This equation is complex and also in the different values $ z $ and $ \theta $, the shape $ f(r) $ is different, so there is not an analytical expression of the jet-quenching parameter in terms of the dynamical exponent and hyperscaling violation parameter. Thus, in figure \ref{www} we plot numerically the jet-quenching parameter as a function of temperature of the plasma at the different values $ z $ and $ \theta $. We see clearly at a fixed temperature, the jet-quenching parameter decreases as the dynamical exponent $ z $ and the hyperscaling violation parameter $ \theta $ increase. These results are in accordance with \cite{mj}.

Based on the results of RHIC, after the collision, the jet quenching parameter decreases temporally as temperature goes down during expansion of QGP. The obtained values for the jet quenching parameter from RHIC is \^{q} $\simeq 5-15~ GeV^{2}/fm $ \cite{ba,esk,dain,jet,jet2}. Comparing these result with ours, we can see that in the range of $ 0.16 \lesssim T \lesssim 0.45 ~GeV$, our results are in consistent with RHIC. 

Interestingly, in \cite{hesh} authors showed that $ z $ and $ \theta $ have different effects on the jet quenching parameter. By increasing $ z $ and $ \theta $, \^{q} decreases and increases, respectively. But we obtained that by increasing $ z $ and $ \theta $, the jet quenching parameter decreases. Thus, one can conclude that the effect of the dynamical exponent $ z $ is in agreement, but the effect of the hyperscaling violation $ \theta $ is different. This is due to the different metric backgrounds and also the difference at the considered ranges of $ z $ and $ \theta $ \cite{mj}.

In the spatial case $ z=1 $, $ \theta=0 $ and $ Q=1 $, one can obtain the following result,
\begin{equation}
\widehat{q}_{SYM}=\sqrt{\lambda} \pi^{3/2} T^{3}\dfrac{\Gamma(\frac{3}{4})}{\Gamma(\frac{5}{4})},
\end{equation}
where is in accordance with \cite{liu}.

\section{Adding a constant electromagnetic field to the background}\label{sec4}
In the previous section, we calculated the jet quenching parameter in the hyperscaling violation metric background. Here, we want to add a constant electromagnetic field on the D-brane and investigate the effect of it on the jet quenching parameter using the method suggested in \cite{field}. This constant B-field that is $ B=Edt\wedge dx_{1}+\mathcal{H}dx_{1}\wedge dx_{2} $, couples to line element (\ref{ads}) where $ E $ and $ \mathcal{H} $ are the NS-NS antisymmetric electric and magnetic fields along the $ x_{1} $ and $ x_{2} $ directions and other components of B-field are zero. Due to that only the field strength is involved in the equation of motion, this ansatz is a good solution to supergravity and the minimal setup to study the B-field correction. 

In the first case, we consider that only the constant electric field turns on, i.e. $ \mathcal{H}=0 $. As before, we should use the light-cone coordinates, $ x^{\pm}=(t\pm x^{1})/\sqrt{2} $, and the static gauge $ \tau=x^{-}, \sigma=x^{2} $. By adding an electric field to the metric background (\ref{ads}), one can find the string action as the following equation,
\begin{equation}
S=\dfrac{1}{2\pi \alpha^{\prime}}\int d\tau d\sigma\sqrt{det(g+b)_{\alpha\beta}}.\label{ng}
\end{equation}
Then one can find the induced b-field and metric on the string worldsheet as following,
\begin{equation}
(g+b)_{\alpha\beta}=r^{2(\alpha +1)}\left( 
\begin{array}{ccc}
\dfrac{1}{2} \left( 1-r^{2(z-1)}f(r)-\dfrac{E}{r^{2(\alpha +1)}}\right)  & 0 & \\
0 &1+\dfrac{r^{\prime 2}}{r^{4}f(r)}  &
\end{array} \right).
\end{equation}
Therefore, the string action (\ref{ng}) becomes,
\begin{equation}
S=\dfrac{L^{-}}{\sqrt{2}\pi\alpha^{\prime}} \int _{0}^{L/2} d\sigma \sqrt{r^{4(\alpha+1)}\left( 1+\dfrac{r^{\prime 2}}{r^{4}f} \right) \left(1- r^{2(z-1)}f-\dfrac{E}{r^{2(\alpha+1)}} \right) }.\label{13}
\end{equation}
\begin{figure}
\begin{center}$
\begin{array}{cc}
\includegraphics[width=74 mm]{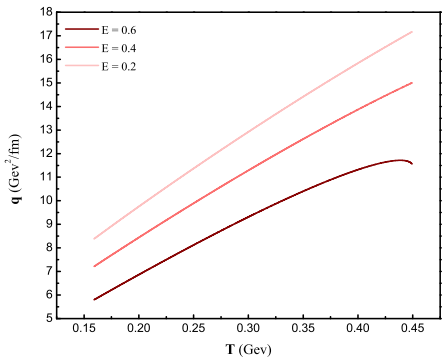}
\end{array}$
\end{center}
\caption{Jet-quenching versus temperature in different $ E $ at $ z=2 $, $ \theta =d-1 $ and $ Q=1 $.}
\label{dd}
\end{figure}
Because the Lagrangian has no $ \sigma $-dependence explicitly, the corresponding Hamiltonian $ H $ is a constant of motion, therefore the equation of motion for $ r $ can be obtain as,
\begin{equation}
r^{\prime 2}=\dfrac{r^{4}f}{2H^{2}}\left( r^{4(\alpha +1)}\left( 1-r^{2(z-1)}f-\eta \right)-2H^{2}  \right),~~~~~\eta:= \dfrac{E}{r^{2(\alpha +1)}}. \label{12}
\end{equation}
Similar to the previous section, we consider the small $ H $ case. Substituting (\ref{12}) into (\ref{13}) in the $ H\rightarrow 0 $ limit, we find,
\begin{equation}
S=\dfrac{L^{-}}{\sqrt{2}\pi\alpha^{\prime}} \int dr \sqrt{\dfrac{r^{4\alpha}}{f} \Big( 1-r^{2(z-1)}f-\eta \Big)}\left( 1+\dfrac{H^{2}}{r^{4(\alpha +1)}\left( 1-r^{2(z-1)}f-\eta \right) }\right),
\end{equation}
and finally, we conclude the following jet quenching parameter,
\begin{equation}
\hat{q}=\dfrac{1}{\pi\alpha^{\prime}}\left( \int_{r_{h}}^{r_{F}} \dfrac{dr}{\sqrt{r^{4(\alpha+2)}f(r)\left( 1-r^{2(z-1)}f(r)-\eta \right)}} \right)^{-1},
\end{equation}
which under defined transformation $ u=r_{h}/r $ is,
\begin{equation}
\hat{q}=\dfrac{1}{\pi\alpha^{\prime}}\Bigg[ \int _{\epsilon}^{1}\dfrac{du}{u^{2}}\dfrac{r_{h}}{\sqrt{\left (\dfrac{r_{h}}{u}\right )^{4(\alpha +2)}f(u)\left(1- \left (\dfrac{r_{h}}{u}\right )^{2(z-1)}f(u)-\eta \right) }} \Bigg]^{-1}.
\end{equation}

Now, we plot numerically the jet quenching parameter versus the temperature in figure \ref{dd}. We see obviously that increasing the electric field leads to decreasing the jet quenching parameter at a fixed temperature which is in agreement with \cite{mj} since the drag force and the jet quenching behave similarly.

In the second case, we consider the constant magnetic field turns on and $ E=0 $. As before, we use the light-cone coordinates, $ x^{\pm}=(t\pm x^{1})/\sqrt{2} $, and the static gauge $ \tau=x^{-}, \sigma=x^{2} $. By adding an magnetic field to the background, we obtain the following induced $ b $-field and metric on the string worldsheet,
\begin{equation}
(g+b)_{\alpha\beta}=\left( 
\begin{array}{ccc}
\dfrac{r^{2\alpha +2}}{2} \left( 1-r^{2z-2}f(r)\right)  & -\dfrac{\mathcal{H}}{\sqrt{2}} & \\
\dfrac{\mathcal{H}}{\sqrt{2}} &r^{2\alpha +2} \left( 1+\dfrac{r^{\prime 2}}{r^{4}f(r)}\right)  &
\end{array} \right).
\end{equation}
Then, the string action (\ref{ng}) becomes,
\begin{equation}
S=\dfrac{L^{-}}{\sqrt{2}\pi \alpha'}\int_{0}^{L/2} d\sigma \sqrt{r^{4\alpha +4}\left( 1-r^{2z-2}f \right)\left( 1+\dfrac{r^{'2}}{r^{4}f} \right) +\mathcal{H}^{2} }
\end{equation}
The Lagrangian is not $ \sigma $-dependence and there is a constant of motion, $ H $. At the end, by performing the similar calculation in the $ H\rightarrow 0 $ limit as before, we obtain the jet quenching parameter as following,
\begin{equation}
\hat{q}=\dfrac{1}{\pi\alpha^{\prime}}\left( \int_{r_{h}}^{r_{F}} dr\sqrt{\dfrac{r^{4\alpha}\left( 1-r^{2z-2}f(r) \right) } {f(r)\Big( r^{4\alpha+ 4}\left( 1-r^{2z-2}f(r)\right) + \mathcal{H}^{2} \Big)^{2}}} \right)^{-1},
\end{equation}
which become with definition $ u=r_{h}/r $ as follow,
\begin{equation}
\hat{q}=\dfrac{1}{\pi\alpha^{\prime}}\left( \int_{\epsilon}^{1} \dfrac{du}{u^{2}}r_{h}\sqrt{\dfrac{\left (\dfrac{r_{h}}{u}\right )^{4\alpha}\left( 1-\left (\dfrac{r_{h}}{u}\right )^{2z-2}f(u) \right) } {f(u)\Big[ \left (\dfrac{r_{h}}{u}\right )^{4\alpha+ 4}\left( 1-\left (\dfrac{r_{h}}{u}\right )^{2z-2}f(u)\right) + \mathcal{H}^{2} \Big]^{2}}} \right)^{-1} .
\end{equation}

\begin{figure}
\begin{center}$
\begin{array}{cc}
\includegraphics[width=74 mm]{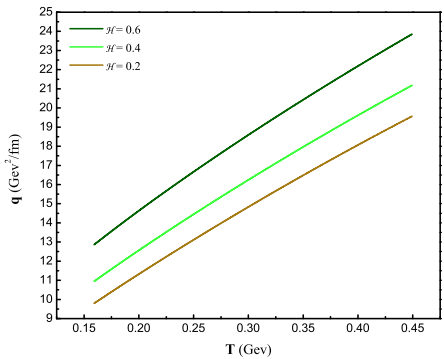}
\end{array}$
\end{center}
\caption{Jet-quenching versus temperature in different $ \mathcal{H} $ at $ z=2 $, $ \theta =d-1 $ and $ Q=1 $.}
\label{db}
\end{figure}

Here, we depict the jet quenching parameter versus the temperature numerically in figure \ref{db}. As we see, increasing the magnetic field leads to increasing the jet quenching. One can obtain the electric and the magnetic fields have the different effects on the jet quenching.

\section{Conclusion}\label{sec5}
In this paper, we have considered a quark-antiquark pair in a non-relativistic background. This background drives from the Einstein-Maxwell-Dilaton theory that consists both scalar and Abelian gauge fields and has two dynamical and hyperscaling violation parameters.

We have investigated the heavy quark potential and the jet quenching parameter in various values $ z $ and $ \theta $ and studied effects of $ T $ and $ Q $. It is obvious that the dynamical and hyperscaling violation parameters have different effects on the potential. Increasing $ z $ and $ \theta $ lead to increasing and decreasing the heavy quark potential, respectively. This results are consistent with \cite{zii}. Also, the potential decreases and increases by increasing $ Q $ and $ T $, respectively. By comparing plots with the conformal case, we have concluded that the potential behaves similar to the conformal case and the closest case to the conformal state is $ z=2 $, $ \theta=d-1 $, $ Q=1 $ and $ r_{h}=1.2 $. Also, the effects of $ z $ and $ \theta $ have been investigated on the jet quenching parameter. It is seen that by increasing $ z $ and $ \theta $, the jet quenching parameter decreases that is in agreement with \cite{mj}. These results are inaccordance with \cite{hesh}, due to the different metric backgrounds and the considered ranges of $ z $ and $ \theta $. Also, the values of the jet quenching that we have obtained are $ 8\lesssim $ \^{q} $ \lesssim 15~ GeV^{2}/fm $, that are consistent with the results of RHIC. 

In section \ref{sec4}, we have added a constant electromagnetic field on the D-brane. We have observed that the jet quenching parameter behaves differently by changing the electric and magnetic fields. Increasing the electric field leads to decreasing the jet quenching parameter. Since the behaviour of the drag force and the jet quenching is similar, this result is in agreement with \cite{mj} and interestingly we have found that the jet quenching increases as $ \mathcal{H} $ increases.

\end{document}